\documentclass{moriond}

\usepackage{amsmath}
\usepackage{amssymb}
\usepackage{amsthm}

\newcommand{\pd}{\partial}
\newcommand\m{\mu}

\newcommand\n{\nu}

\renewcommand\a{\alpha}
\renewcommand\b{\beta}
\renewcommand\l{\lambda}

\def\be{\begin{equation}}
\def\ee{\end{equation}}
\def\bea{\begin{eqnarray}}
\def\eea{\end{eqnarray}}

\newcommand{\Photo}{}

\begin{document}
{\normalsize
\flushright CERN-PH-TH-2015-143 \\}
\vspace*{-1cm}

\vspace*{4cm}
\title{Lorentz violation in gravity}

\author{Diego Blas}

\address{CERN, Theory Division, 1211 Geneva, Switzerland.}

\maketitle\abstracts{
The study of gravitational theories without Lorentz invariance  plays an important role to understand  different
aspects of gravitation. In this short contribution we will describe the construction, 
main advantages and some phenomenological considerations associated with the presence of a preferred time direction.}

\section{Introduction}

One hundred years after its formulation, General Relativity (GR) is living a golden era of continuous verifications of its predictions, at many scales and in very different processes \cite{Will:2014bqa,Berti:2015itd,Clifton:2011jh}. The agreement of data with 
GR predictions is both astonishing (given the range of scales probed) and frustrating (since GR can not be a complete quantum theory, 
but we lack experimental guidance towards its completion)\footnote{The existence of dark matter and dark energy is sometimes considered as a hint towards the construction of alternatives to GR. This motivation is certainly valid though the standard paradigm based on GR is consistent both theoretically and phenomenologically.}. 

  Besides confirming the predictions of GR, the current data can also be used to constrain possible deviations. This is an important program from which we
  can learn  about the robustness of the different properties of GR,  the benefits of modifying them 
and the viability of the resulting alternative theories. This brief note is devoted to a particular modification suggested by  theories of gravity with a better quantum behaviour than GR. To attain this, they abandon one of the principles of GR: Lorentz invariance. 
We will only discuss the case of a preferred frame defining a time direction at every point of space-time. After introducing the formalism and 
explaining its possible relation to quantum gravity, we will proceed to derive some of its phenomenological consequences. 
For further information, the reader can consult the
recent review paper \cite{Blas:2014aca}.

\section{Theoretical construction}\label{sec:theo}

We will consider metric theories where there is a local preferred time direction represented by a time-like vector field $u^\mu$ satisfying
\be
\label{eq:unit}
u_\mu u^\mu=1.
\ee
If this vector is otherwise generic, these theories are known as Einstein-aether theories \cite{Jacobson:2000xp}. Their relation to quantum gravity is not completely clear, but their study
may be important for  
 fundamental theories of gravity where Lorentz invariance is broken (spontaneously or fundamentally) by the selection of a preferred frame. A more concrete example of how this may happen is provided by Ho\v rava gravity, which 
assumes the existence of a preferred foliation of space-time into space-like hypersurfaces \cite{Horava:2009uw}. 
This allows to construct a theory of gravity renormalizable by power-counting and close to GR at low-energies. In terms of the vector satisfying (\ref{eq:unit}), the theory requires the existence of a field $\varphi$ representing the foliation and 
\be
\label{eq:kh}
u_\m\equiv\frac{\pd_\mu \varphi}{\sqrt{g^{\m\n}\pd_\m  \varphi\pd_\n \varphi}}.
\ee
The generic theories defined with a vector of the form (\ref{eq:kh}) are known as khronometric theories \cite{Blas:2010hb}.
To construct the action we write the different operators including $u^\mu$ and $g^{\m\n}$, covariant under diffeomorphisms and organized in a derivative expansion  (we also assume CPT),
\be 
\label{ae-action} 
S = -\frac{M_0^2}{2}\int  ~d^{4}x \sqrt{-g}~ \left(R
+K^{\a\b}{}_{\m\n} \nabla_\a u^\m \nabla_\b u^\n+\lambda (u^\m u_\m-1)+\frac{{\mathcal O^{n+2}}}{M^n_\star}\right)\,,
\ee 
where $g$ and $R$ are the metric determinant and the Ricci scalar and 
\be
\label{eq:K}
K^{\a\b}{}_{\m\n} \equiv c_1 g^{\a\b}g_{\m\n}+c_2\delta^{\a}_{\m}\delta^{\b}_{\n}
+c_3 \delta^{\a}_{\n}\delta^{\b}_{\m}+c_4 u^\a u^\b g_{\m\n}\, ,
\ee 
We use the constant $M_0$ instead of $M_P$ for the mass scale
in front of the Einstein-Hilbert action to distinguish it from the quantity appearing in Newton's law \cite{Blas:2014aca}. 
By the last term in \eqref{ae-action}  we indicate the higher dimensional operators, which we assume to be suppressed by a common scale $M_\star$.
We imposed the restriction (\ref{eq:unit}) through a Lagrange multiplier $\lambda$. In the khronometric
case, this is not necessary. Furthermore, the condition \eqref{eq:kh} implies that one of the terms in (\ref{eq:K})  can be expressed in terms
of the others. One then absorbs the $c_1$ term into the other three terms by multiplying the second, third and forth term respectively by the new couplings
\be\label{eq:EAtoKH}
\lambda\equiv c_2, \quad \beta\equiv c_3+c_1, \quad \alpha\equiv c_4+c_1.
\ee
For a reformulation of \eqref{ae-action} in terms of geometrical quantities of the congruences of  $u^\mu$ see \cite{Jacobson:2013xta}.

\section{Short distance modifications}\label{sec:short}

Let us first discuss  the operators of \eqref{ae-action} suppressed by the scale $M_\star$. 
Since we suppose that  Lorentz invariance is broken in a regime where gravity is  weakly coupled, one can  
start parameterizing the
changes in gravitation by considering linear equations for the perturbations of the metric with Lorentz violating (LV) terms. Assuming that parity and time reversal are not violated and that the  equations
are  at most second order in time derivatives, we can introduce  the dispersion relations
\be
\label{eq:lvgw}
\omega^2=p^2\left(1+\sum_{n=1}^{L}\alpha_n\left(\frac{p}{M^{gw}_{\star}}\right)^{2n}\right),
\ee
for the propagating degrees of freedom (e.g the graviton) and the modified Poisson's equation 
\be
\label{eq:new}
p^2\left(1+\sum_{n=1}^{L}\beta_n\left(\frac{p}{M^{\phi}_{\star}}\right)^{2n}\right)\phi=-\frac{1}{2M^2_0}\tau_{00},
\ee
for the potentials $\phi$  sourced by matter's energy, represented by $\tau_{00}$.
The mass scales $M_{\star}^\phi$  and $M_{\star}^{gw}$ are kept independent, even if they are both related to $M_\star$. 

Let us first discuss the modifications of the graviton's dispersion relation, Eq.~\eqref{eq:lvgw}. 
If the gravitational waves (GW) have  the dispersion relation \eqref{eq:lvgw}, this modifies the frequency dependence in the 
 propagation of the wave-fronts, which may be observed by future detectors of GW. 
These effects will be very suppressed  if we assume that $M^{gw}_{\star}\approx M^{\phi}_{\star}$, given that the latter are constrained to be  $M^{\phi}_{\star}< (\mu m)^{-1}\approx 10^{-2} \ \mathrm{eV}$ (see below).
They may still have an impact for the GWs generated in the primordial universe since in this case the typical energies during production may be almost as high as $M_P$. Thus, if  primordial GWs are observed, the range of energies  at which LV is tested  (in fact any short-distance modification) will improve dramatically.

More is known about possible deviations of the potentials at high-energies, Eq.~\eqref{eq:new}. Let us focus on the case relevant for 
the short distance behaviour of Ho\v rava gravity 
where only  $\beta_2\neq 0$, and absorb its value into $M_{\star}^\phi$ ($M_{\star}^\phi\mapsto  M_{\star}^\phi \beta_2^{1/4}2^{1/8})$. 
Taking a point particle of mass $m_{pp}$ at rest as a source, the solution of Eq.~\eqref{eq:new} away from the source is
\be
\label{eq:short}
\phi=-\frac{m_{pp}}{8\pi M_0^2 r}\left[1-e^{-M_{\star}^\phi r}\cos\left(M_{\star}^\phi r\right)\right],
\ee
where $r$ is the distance from the source. 
This potential regularises the divergent behaviour of Newton's potential at small distances. Furthermore,
at scales where the deviations start to be important,  it is similar
to the potentials that have been considered  to constrain the deviations from Newton's law at short distances \cite{Kapner:2006si}
\be
\label{eq:shortpot}
\phi=-\frac{m_{pp}}{8\pi M_0^2 r}\left[1+\tilde\alpha\, e^{-r/\tilde \lambda}\right].
\ee
From these works, one concludes that a bound of the form $M_{\star}^\phi \lesssim (\mu m)^{-1}$ should apply.  
However, the differences between the potentials of Eq.~\eqref{eq:shortpot} and \eqref{eq:short} are important, e.g. the potentials  in \eqref{eq:shortpot} are singular at short distances except for $\tilde\alpha=-1$  and they do not present the oscillatory behaviour 
of \eqref{eq:short}. Thus,  a precise bound on $M_{\star}^\phi$ requires a reanalysis of the experimental data.

Finally, even though our previous discussion was organised around the modified equations \eqref{eq:lvgw} and 
\eqref{eq:new},  the effects of LV at short distances (high energies)  may also be important 
for the  background evolution in the primordial universe, see  \cite{Blas:2014aca} for the relevant references.

\section{Long distance modifications}\label{sec:long}

By long distance modifications we mean the theory whose gravitational action is given by \eqref{ae-action} in the limit $M_\star \to \infty$.
Independently of the coupling to matter, some bounds can be derived on the constants $c_i$ from 
stability of perturbations around a Minkowski background and requiring the absence of gravitational Cherenkov radiation\cite{Blas:2014aca}. To find the phenomenological predictions of the theory we first need to
understand how $u^\m$ and $g_{\m\n}$ couple to matter. In principle, any coupling should be allowed. The generic consequence would be the presence of big deviations from Lorentz invariance also in the standard model of particle physics.  These deviations are extremely small, cf. \cite{Liberati:2013xla}, which makes it natural to assume that, as long as gravitational test are concerned, matter
is not coupled to $u^\m$ at all. Explaining how this can happen in a natural way while keeping the other couplings to $u^\m$ 
not too small remains an open challenge for the set-up. Different possibilities have been explored  \cite{Blas:2014aca}, but no definite mechanism
has been produced yet.  Notice that for dark matter and dark energy one can keep these couplings arbitrary. 

Bounds on the LV parameters in \eqref{ae-action} come from different observations. Assuming that the preferred frame $u^\mu$  is aligned with 
the CMB, which is a reasonable supposition \cite{Will:2014bqa,Blas:2014aca},  one finds
that the gravitational potential in the Solar System is modified by the presence of two post-Newtonian parameters $\alpha_1$ and $\alpha_2$. These two parameters are functions of the LV parameters. For instance, for the
khronometric parameters \eqref{eq:EAtoKH} they read
\be
\alpha_1=-4(\alpha-2\beta), \quad \alpha_2=\frac{(\alpha-2\b)(\a-\l-3\b)}{2(\lambda+\beta)}.
\ee
Current observations yield the bounds\cite{Will:2014bqa} $|\alpha_1|<10^{-4}$ and $|\alpha_2|<10^{-7}$.  

Further bounds can be obtained from strongly gravitating bodies. In this case, even if  matter is not coupled to $u^\m$,   the gravitons 
in the body are, and for objects with large gravitational fields (very compact) the body as a whole will   {\it effectively} feel the presence of 
$u^\m$. To parametrize this for the phenomena at large distances with respect to the size of the source, one can assume the action of the 
point particle to be   
\be
\label{eq:acsens}
S_{\rm pp \; A}=-\int ds_A \tilde m_A(\gamma_A), 
\ee
where $\tilde m_A$ is a function of   $\gamma_A\equiv u_\m v_A^\m$ and  $v_A^\m$ is the four-velocity of the source (see
\cite{Damour:1992we} for similar descriptions in scalar-tensor theories). Finally, $ds_A$ is the line element of the trajectory.  If $\gamma_A \ll 1$, one can expand the action to second order in $\gamma_A$ and describe the physics in terms of\begin{align}
\label{sigma-def}
\tilde m_A\big|_{\gamma_{A} = 1}, \quad \sigma_{A} &\equiv - \left.\frac{d \ln \tilde{m}_{A}(\gamma_A)}{d \ln \gamma_{A}}
\right|_{\gamma_{A} = 1}\,.
\end{align}
The parameters $\sigma_A$ are called \emph{sensitivities} and  represent the effective coupling of the source to $u^\m$. 
The presence of these couplings introduces an extra force in the dynamics of binary systems, which precludes the conservation of the usual momentum sourcing gravitational waves\footnote{There is
still a conserved momentum associated with translation invariance, but it differs from the one of GR.}. This implies the emission of dipolar radiation, which is absent in GR. Since the latter is enhanced with respect to the quadrupolar
radiation by a factors $(c/v)^2$, where $v$ is the typical orbital velocity of the system, this means that even for very small $\sigma_A$ (corresponding to neutron stars), one can get very tight bounds on the LV parameter
space by observing the radiation damping of binary pulsars \cite{Yagi:2013ava}. The same is also true  for solitary pulsars, where the bounds come from changes in the spin-precession. For these bounds to relate
to the fundamental parameters in \eqref{ae-action} one needs to compute the numbers $\sigma_A$ for different sources, which was done in \cite{Yagi:2013ava}.

Finally,  cosmological observations produce further bounds on the LV parameters. Remarkably, these also apply to the possible LV in the dark matter. The bounds come from different
observations: first, the Friedmann equation is modified by a renormalization of Newton's constant depending on the LV parameters. This deviation can be constrained with the data from big bang nucleosynthesis, by the growth of structure (controlled by the local $G_N$) and CMB observations \cite{Audren:2014hza}. Furthermore, the existence of the vector field $u^\m$ introduces a source of anisotropic stress present at many scales, and 
which can be constrained
by CMB observations. Similarly, the possible coupling of dark matter to $u^\m$ introduces an extra force in dark matter, which may violate the weak equivalence principle. This has consequences
for the gravitational dynamics of dark matter \cite{Audren:2014hza}. The cosmological observations from the regimes where linearized cosmology is applicable imply bounds close to the percent level for
all the previous couplings \cite{Audren:2014hza}. These bounds will improve once the consequences for non-linear scales (scales below  $10$ Mpc) are understood.


\end{document}